\documentclass{article}

\usepackage{microtype}
\usepackage{graphicx}
\usepackage{subfigure}
\usepackage{booktabs} 
\usepackage{csvsimple}
\usepackage{hyperref}

\usepackage[accepted]{icml2025}

\usepackage{amsmath}
\usepackage{amssymb}
\usepackage{mathtools}
\usepackage{amsthm}

\usepackage[capitalize,noabbrev]{cleveref}

\theoremstyle{plain}

\theoremstyle{definition}

\theoremstyle{remark}

\usepackage[textsize=tiny]{todonotes}

\icmltitlerunning{Scaling and Data Saturation in Protein Language Models}

\begin{document}

\twocolumn[
\icmltitle{Scaling and Data Saturation in Protein Language Models}



\icmlsetsymbol{equal}{*}

\begin{icmlauthorlist}
\icmlauthor{Aviv Spinner}{align}
\icmlauthor{Erika DeBenedictis}{align}
\icmlauthor{Corey M. Hudson}{align}

\end{icmlauthorlist}

\icmlaffiliation{align}{The Align Foundation}

\icmlcorrespondingauthor{Aviv Spinner}{aviv@alignbio.org}
\icmlcorrespondingauthor{Corey M. Hudson}{corey@alignbio.org}

\icmlkeywords{Machine Learning, Biology, Scaling Laws}

\vskip 0.3in
]



\printAffiliationsAndNotice{}  

\begin{abstract}
Data in biology is redundant, noisy, and sparse. How does the type and scale of available data impact model performance? In this work, we specifically investigate how protein language models (pLMs) scale with increasing pretraining data. We investigate this relationship by measuring the performance of protein function prediction on a suite of pLMs pretrained on yearly snapshots of UniRef100 from 2011 to 2024. We find no evidence of model saturation on this task: performance improves---but not monotonically---with added data, and this trend differs between unsupervised and supervised experiments. Using a well-characterized \(\beta\)-Lactamase protein from \textit{E. coli}, we find that unsupervised model predictions get better year-over-year, though they do not yet consistently perform better than the supervised baseline. Our results underscore the need for targeted data acquisition and deeper study of data scaling in protein modeling. All training, inference, analysis, and visualization code is available at: \url{https://github.com/Align-to-Innovate/data-saturation-and-scaling}.

\end{abstract}

\section{Introduction}
\label{intro}

Protein fitness prediction and design is in a period of explosive growth. The successful application of large language models (LLMs) to biological problems has spurred the development of powerful tools. While scaling laws related to compute resources, model parameters, and data quantity have become well-established in the field of natural language processing (NLP) \cite{kaplan2020scaling, hernandez2021scaling, hoffmann2022training}, recent work on probing scaling laws on downstream tasks demonstrates inconsistency \cite{lourie2025scaling}. Lourie et al. conduct a meta-analysis across 46 tasks and find that only 39\% demonstrate predictable scaling behavior, with the remainder exhibiting nonmonotonic, inverse, or trendless scaling. These findings challenge assumptions that pretraining loss reliably predicts downstream performance and emphasize the need to understand the specific conditions under which scaling laws hold. Further, analogous studies of scaling laws around \textbf{biological data} remain largely uncharacterized. Although the field is beginning to explore scaling in terms of biological systems \cite{elnaggar2021prottrans, hesslow2022rita, cheng2024training, fournier2024protein, li2024feature, nguyen2024sequence}, to our knowledge, there is currently no comprehensive study that investigates how scaling laws relate to data scaling for the task of zero-shot and semi-supervised protein fitness prediction.

\subsection{Biological Data}

The growth of biological data over the past two decades has been extraordinary. This has enabled the biological machine learning community to answer difficult questions about the interplay of data scale, model performance and how these relate to challenging tasks in biological prediction. Much of this growth in biological data has been in the form of massive sequencing databases such as Uniprot \cite{uniprot}, MGnify \cite{mgnify}, OMG \cite{omg}, and the Big Fantastic Database (BFD)\cite{jumper2021highly}. Uniprot is the oldest of these and has experienced considerable growth over its lifetime (Figure \ref{conceptfig}B) concomitant with improvements in sequencing. MGnify has also provided considerable growth in sequences, through the incorporation of billions of non-redundant metagenomic assembly sequences. These databases have provided the biological machine learning community with fantastic opportunities to grow the field and expand it into previously impossible questions, evidenced by all of the models trained on these collection of sequences \cite{jumper2021highly, omg, notin2022tranception, lin2023evolutionary, madani2023large}. Despite the dramatic growth of protein sequence databases, current sequencing efforts capture only a tiny fraction of nature’s true protein diversity. Estimates suggest that Earth harbors up to $10^{12}$ microbial species, most of which remain unsequenced and many of which possess proteins of previously unexplored functions, highlighting how little of the protein universe has been incorporated into AI models \cite{louca2019census}. This gap in sampling and sequencing space represents a potentially fundamental challenge in model sufficiency and protein fitness prediction.

On the other hand, several highly curated data repositories exist that provide annotations and experimental measurements of protein mutational fitness data \cite{rubin2025mavedb}. Resources such as ProteinGym \cite{notin2023proteingym} and the TAPE benchmark \cite{rao2019evaluating} have become standard for evaluating machine learning models on tasks like mutation effect prediction and transfer learning. Testing model performance on experimental data is necessary, and still it only describes a narrow slice of the protein universe and cannot fully capture the breadth, complexity, and noisiness of biological data encountered in real-world applications. 


\subsubsection{Challenges}

However, more data does not necessarily equate to better AI model performance. Biological data presents several unique challenges, only some of which are improved by increased data abundance:
\begin{itemize}
    \item \textbf{Redundancy and imbalance}: Overrepresentation of specific families or taxa can bias training and obscure generalization \cite{ding2024protein, Poux2016}.
    \item \textbf{Annotation sparsity}: Many sequences lack experimental validation or consistent functional labels \cite{RAUER2021108}.
    \item \textbf{Noisy and heterogeneous data sources}: Sequences originate from a mix of high-throughput and manual pipelines, often with varied quality standards \cite{Chorlton2024_reference_issues}.
    \item \textbf{Functional ambiguity}: Proteins can have multiple or context-dependent functions, making supervised learning difficult \cite{Jeffery2023_ProteinFunctionChallenges}.
\end{itemize}

In spite of these data challenges, there has been consistent growth in the applications of protein language models (pLMs) across a variety of tasks. However, pLMs are not fully task agnostic. More complex protein tasks (e.g., moonlighting proteins, experimental outcomes, protein fitness and function, etc.) put a greater burden on the model's underlying ability to generalize \cite{Zhou2024}. In practice, increasing the complexity of biological tasks demands more richly labeled datasets. 

\subsection{Scaling Laws}
Scaling laws have helped researchers balance resources between compute, parameters, and training data in order to make optimal models. Teams have measured scaling laws of parameters and compute for protein language models \cite{elnaggar2021prottrans, cheng2024training, fournier2024protein, li2024feature, serrano2024protein} as well as published scaling laws experiments as new models are released \cite{notin2022tranception, bhatnagar2025scaling, rives2021biological}.

\begin{figure}[ht]
\vskip 0.2in
\begin{center}
\centerline{\includegraphics[width=\columnwidth]{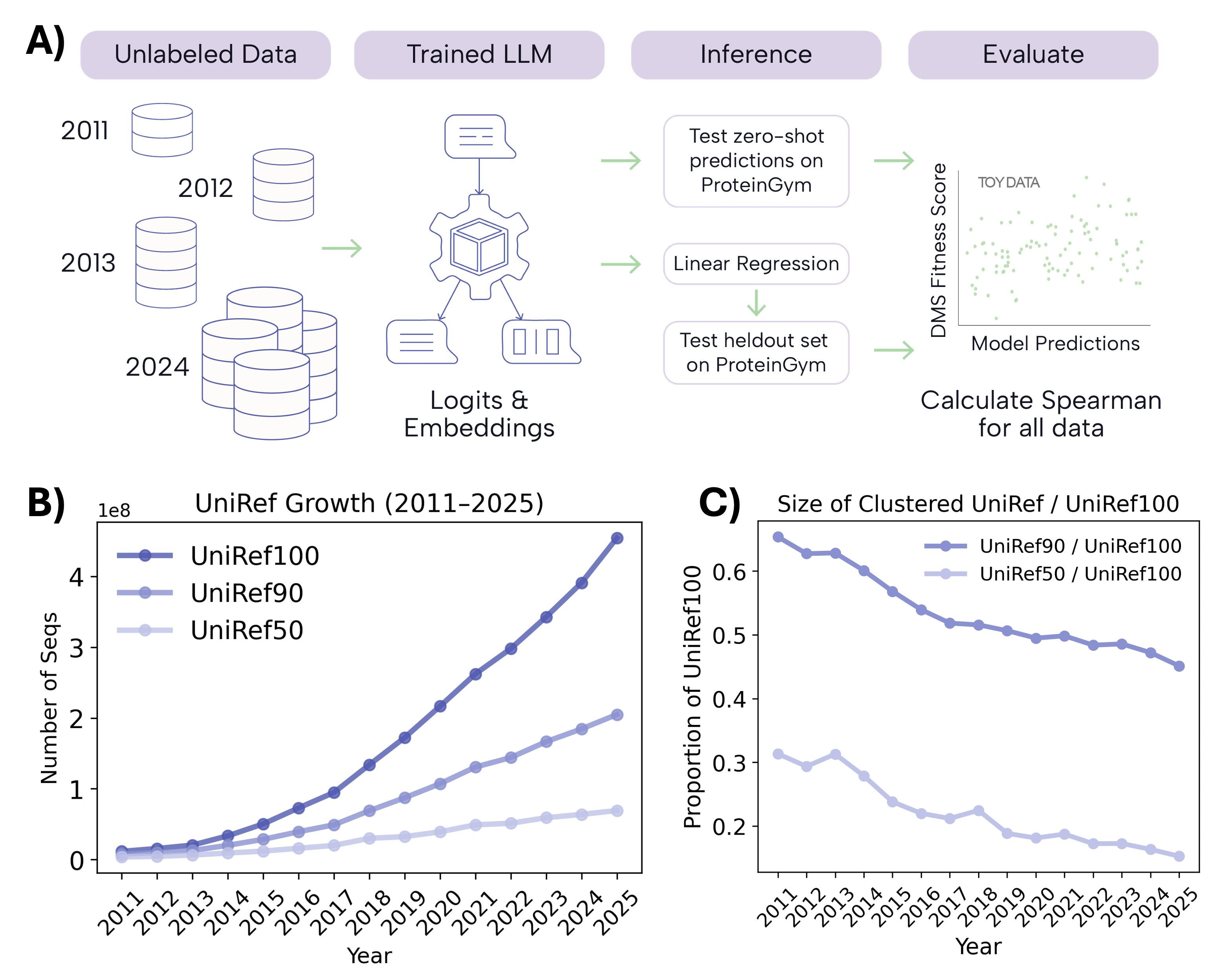}}
\caption{A) \textbf{Overview of the benchmarking pipeline: protein sequences from UniRef train a suite of protein language models.} We use these models to compute logits and embeddings for each sequence in ProteinGym. These are used for zero-shot prediction or representations for linear regression, respectively. Model performance is assessed using Spearman correlation with experimental fitness scores.
B) \textbf{Growth of UniRef clusters from 2011 to 2025.} UniRef100 increases fastest, followed by UniRef90 and UniRef50, indicating increasing redundancy in the dataset over time. See Appendix Table \ref{tab:uniref-years} for sequence counts.
C) \textbf{Proportion of UniRef100 covered by UniRef90 and UniRef50 over time.} Declining ratios indicate that newly added sequences are increasingly similar to existing entries, reflecting decreasing diversity in the dataset.}
\label{conceptfig}
\end{center}
\vskip -0.2in
\end{figure}

Despite this, our understanding of biological data scaling laws remains exceptionally limited. While there is consensus that 'more data is better', this heuristic often obscures the nuanced reality that not all data contributes equally to model performance. As previously noted, high sequence redundancy can degrade model performance while strategic data curation can improve outcomes \cite{omg}. Diminishing returns in performance are frequently observed as datasets grow in size \cite{gordon2024protein}. 

Recent work has begun to address these questions. The AMPLIFY suite of models \cite{fournier2024protein}, trained on UniRef100 snapshots, spanning from 2011-2024, offers a unique window into how model performance changes as the biological sequence pretraining data expands. While many models offer publicly available checkpoints across varying parameter sizes, no other set of models (including ESM\cite{lin2023evolutionary}, ProGen\cite{nijkamp2023progen2}, and other protein language models) provides a systematic investigation of training data effects with checkpoints released across distinct training data splits. Despite their relatively small parameter counts and single-seed training, AMPLIFY models exhibit competitive performance, remarkable speed, and unique training data splits, enabling a new set of questions around data growth and model generalization to be addressed.

\subsection{Our contribution}
Here, we first seek to understand the performance of pLMs trained on increasing amounts of unlabeled, publicly-hosted sequence data. We use the suite of AMPLIFY models trained on time-based snapshots of UniRef100 from 2011-2024, as shown in Figure \ref{conceptfig}A. Notably, we use Spearman correlation of log-likelihoods of ProteinGym sequences and direct experimental measurements of mutant fitness in ProteinGym as our metric for model performance. 

If biological data follows scaling laws similar to those observed in other fields, then model performance should improve predictably as more data is incorporated into the model. In this way, we would project data saturation on the field and provide concrete steps for achieving that. To evaluate the presence or absence of this phenomenon, we use the sequence embeddings from those models in concert with assay data on protein sequences to understand how semi-supervised learning performance changes with both unlabeled training data and labeled training data. In both of these cases, we fail to find the clear hallmarks of scaling laws. From this we infer that for the protein function task, we have not yet reached data saturation.

\section{Methods}

\subsection{Task Datasets}
We focus on protein variant effect prediction using the ProteinGym benchmark \cite{notin2023proteingym}, specifically the substitution datasets from the ProteinGym 1.0 release, with proteins shorter than AMPLIFY’s context limit of 2048 amino acids. The \texttt{DMS\_score} from the resulting 213 datasets is used as the phenotypic label of protein function. We exclude four datasets because of max context lengths of sequences in AMPLIFY: \texttt{A0A140D2T1\_ZIKV\_Sourisseau\_2019}; 9576 variants, \texttt{BRCA2\_HUMAN\_Erwood\_2022\_HEK293T}; 265 variants, \texttt{POLG\_HCVJF\_Qi\_2014}; 1630 variants and \texttt{POLG\_CXB3N\_Mattenberger\_2021}; 15711 variants, in total comprising 1\% of ProteinGym.

\subsection{Models}

\subsubsection{Zero-shot} 

There is only one collection of models, to our knowledge, that share a unified training scheme across many different pretraining datasets. We therefore use the suite of 14 AMPLIFY models \cite{fournier2024protein} trained on yearly releases of UniRef100 from 2011 to 2024. 

\textbf{Sequence Log Probability:}
To compute the zero-shot ``fitness'' of a protein sequence, we follow the standard practice of approximation using the log probability from the model. We apply a softmax over the output logits at each position to extract a probability for each of the actual tokens, and sum their logarithms.





\begin{figure}[ht]
\vskip 0.2in
\begin{center}
\centerline{\includegraphics[width=\columnwidth]{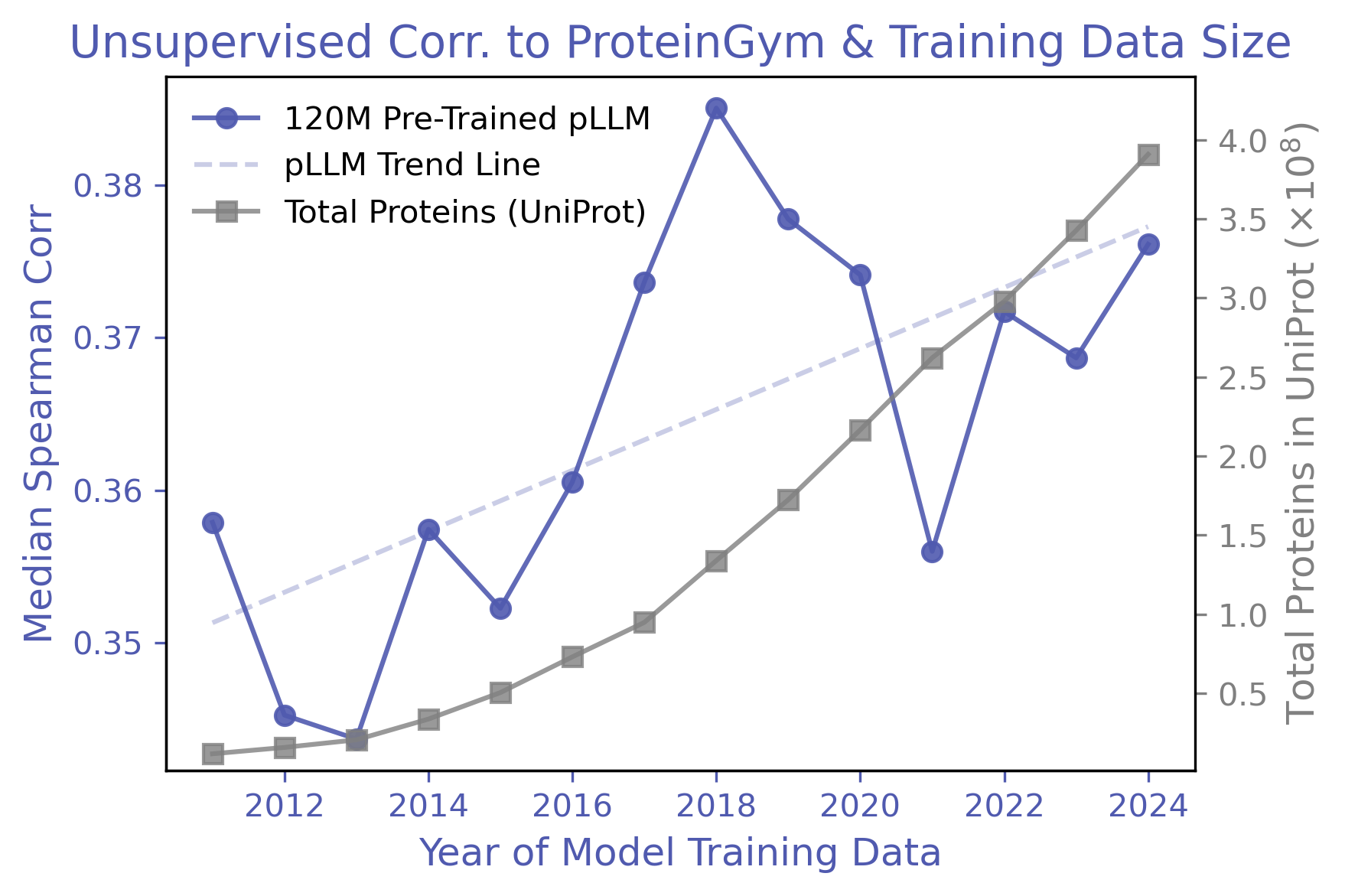}}
\caption{\textbf{Sequence growth does not guarantee performance gains.}
The total number of sequences in UniProt increases steadily from 2011 to 2024. Meanwhile, the Spearman correlation of an AMPLIFY model trained on yearly data sources fluctuates and does not show a monotonic improvement in performance.}
\label{pg_correlation_by_year}
\end{center}
\vskip -0.2in
\end{figure}

To evaluate zero-shot performance, we compare the log-likelihood assigned by each model to the corresponding \texttt{DMS\_score} label for each mutant in ProteinGym, using the Spearman correlation coefficient as the evaluation metric, shown in Figure~\ref{pg_correlation_by_year}.

\subsubsection{Supervised}

\textbf{Sequence Embeddings:}
To obtain each protein's sequence embedding for supervised learning tasks, we tokenize the sequence, pass it through an AMPLIFY model, extract the hidden states from the final layer, and compute the mean over the sequence dimension.



We train ridge regression models \cite{hoerl1970ridge} using sequence embeddings as input features (\(x\)) and DMS fitness values as labels (\(y\)) to study scaling in a supervised learning context. Although more sophisticated semi-supervised and supervised architectures exist \cite{hsu2022learning, notin2023proteinnpt, groth2024kermut}, here we use ridge regression due to its computational efficiency, modularity (i.e., compatibility with diverse inputs and labels), and high interpretability. 
    
\begin{figure}[ht]
\vskip 0.2in
\begin{center}
\centerline{\includegraphics[width=0.8\linewidth]{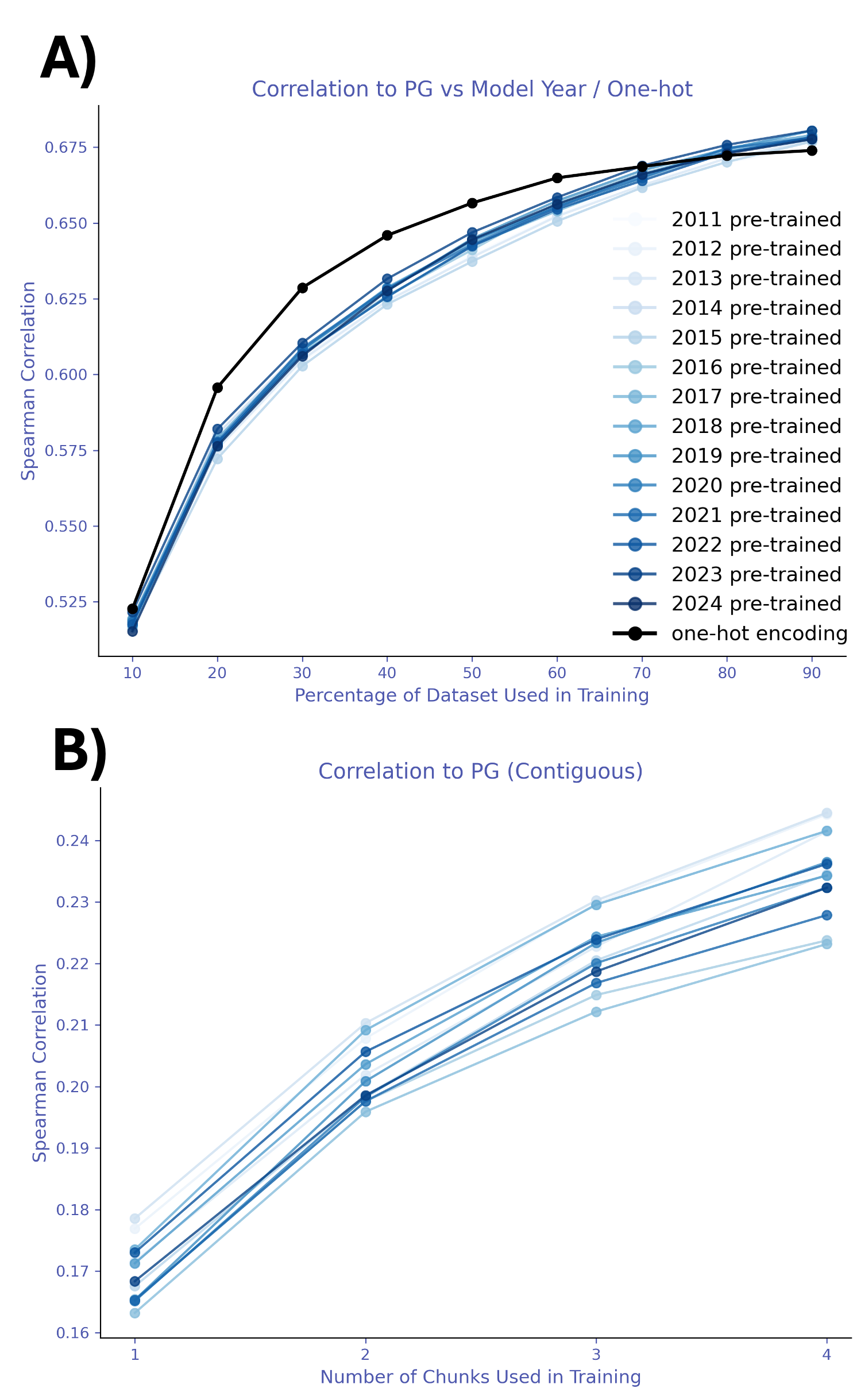}}
\caption{\textbf{Effect of semi-supervised training data size and split strategy on model performance.} \textbf{A) Random train/test splits} with increasing amounts of labeled data show minimal performance differences across AMPLIFY models. One-hot encoding outperforms model embeddings until a large volume of labeled data is used.
\textbf{B) Contiguous train/test splits} reveal no clear relationship between model performance and the amount of pretraining data seen by each AMPLIFY model.}
\label{rand}
\end{center}
\vskip -0.2in
\end{figure}

\textbf{Splitting Within Datasets:} We implement three versions of the cross-validation schemes introduced in Tranception \cite{notin2022tranception} and used in ProteinNPT \cite{notin2023proteinnpt} for benchmarking. In a \textit{Random} split, we randomly partition the data into train/test splits in 10\% increments, ranging from 10/90 to 90/10. Each split ratio is replicated five times with different random seeds (Figure~\ref{rand}A). In the \textit{Contiguous} scheme, we divide each protein sequence into five equal-length contiguous segments and train/test on subsets of these contiguous chunks (Figure~\ref{rand}B). In the \textit{Modulo} scheme, we split the positions of the protein into five groups and conduct train/test splits on subsets of these (Figure~\ref{modulo_fig2b}) 

We conduct experiments over all possible train/test combinations. For example, training on 2 chunks and testing on the remaining 3 yields: $\binom{5}{2} = 10$ combinations. We exclude multi-mutants since they cannot be reliably assigned to a single train or test partition when chunk boundaries are enforced. Results are show in Figure~\ref{rand}B.

Note, the \textit{Random} split can introduce data leakage, as the same mutational position may appear in both training and test sets, allowing the model to memorize positional effects. In contrast, \textit{Contiguous} and \textit{Modulo splits} isolate positions between train and test sets, eliminating this leakage. This has a striking impact on results: for one-hot encodings, random splits yield high Spearman correlations that often exceed those of pretrained embeddings, while in the leakage-controlled splits, performance drops near zero, highlighting the importance of proper evaluation (Figure \ref{modulo_fig2b}).

\begin{figure}[ht]
\vskip 0.2in
\begin{center}
\centerline{\includegraphics[width=\columnwidth]{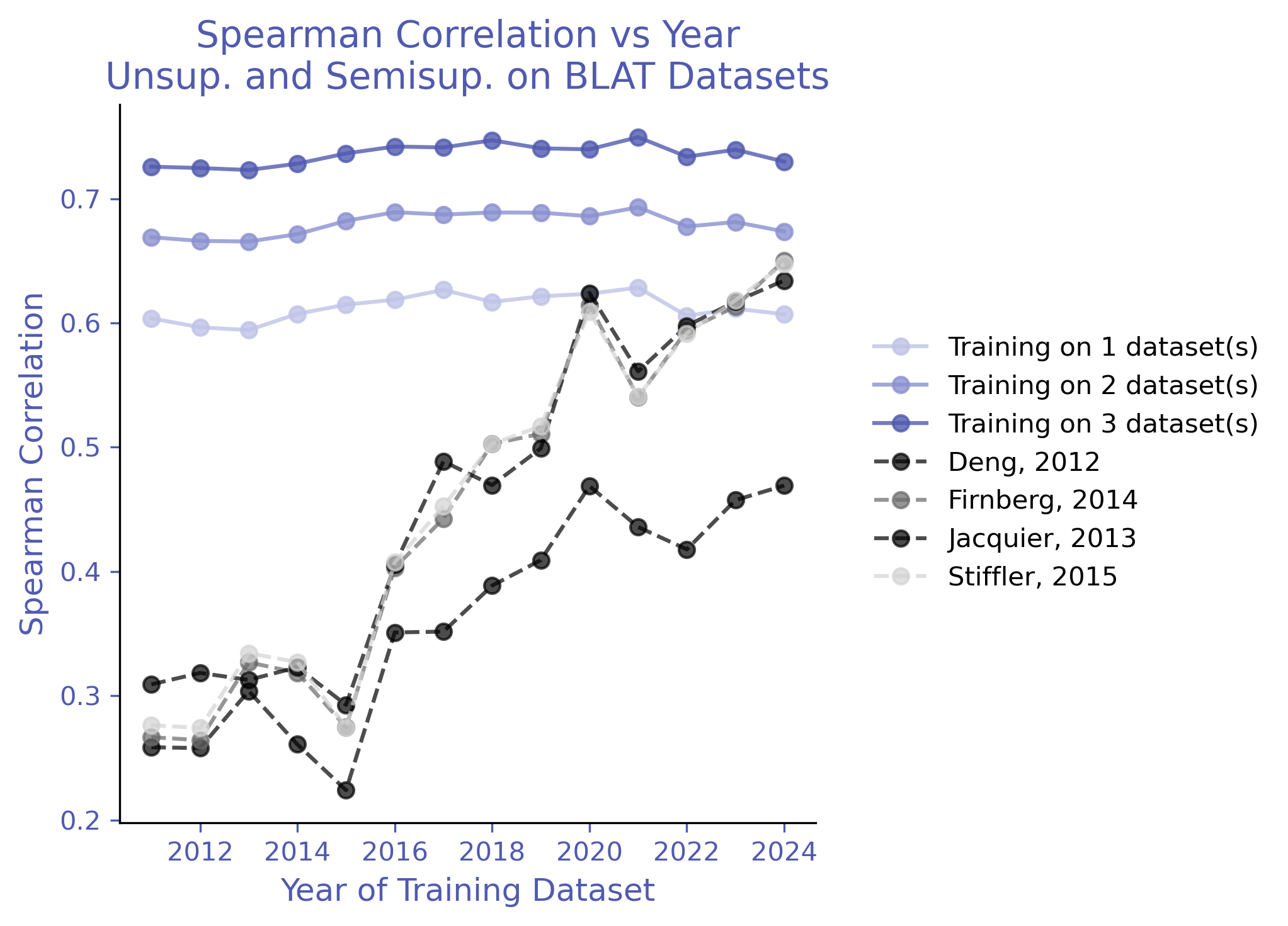}}
\caption{\textbf{Tradeoff between pretrained model and semi-supervised learning.} Unsupervised performance (grey traces) improves as AMPLIFY models are pretrained on more UniRef100 data. In contrast, semi-supervised learning (blue traces) yields relatively stable performance across models, showing little benefit from additional pretraining data.}
\label{BLAT}
\end{center}
\vskip -0.2in
\end{figure}

\textbf{Splitting Between Datasets:} To assess how information from one experiment may generalize to future assays---a key motivation for applying machine learning to protein function prediction---we implement structured cross-dataset evaluation. Within ProteinGym, the \(\beta\)-Lactamase protein includes four deep mutational scanning datasets collected across multiple years (2012–2015), with nearly identical sets of mutations assayed in each. We define a structured split in which models are trained on earlier experiments and evaluated on later ones. Specifically, we train on the 2012 dataset and test on those from 2013, 2014, and 2015; then train on 2012–2013 and test on 2014–2015; and finally train on 2012–2014 and test on 2015 alone. This setup reflects a realistic prospective scenario and enables evaluation of generalization across time and experimental conditions (Figure~\ref{BLAT}).

\section{Results and Discussion}

We find that model performance does not increase monotonically with the amount of sequence data used for training. The Spearman correlation between experimental data and model-predicted fitness fluctuates year-to-year, showing some decreases even with more training data (Figure~\ref{pg_correlation_by_year}). For instance, there is a consistent drop in performance between the years of 2018 to 2021, despite an additional billion sequences (and several hundred billion tokens) in UniRef100. One possible interpretation is that between 2014 and 2018, the sum of data added to the model had more relative influence on the model's predictive capabilities of protein function than the data that were added between 2019 and 2021. Overall, the variability in correlation suggests that the model remains sensitive to the specific sequences added or removed at each timepoint, indicating that it has not yet reached data saturation sufficient for robust generalization.

Further, we split the datasets by two classifications in ProteinGym: \texttt{MSA\_Neff\_L\_category} and \texttt{coarse\_selection\_type}. When stratifying by MSA depth, proteins with larger MSAs (as measured by Neff/L) tended to show improved prediction performance with later model training years, while those with smaller MSAs showed weaker or declining trends over time (Figure~\ref{by_depth_fig1}). Similarly, when partitioning by functional assay type, proteins evaluated using Organismal Fitness as the readout exhibited the most consistent improvement over time, whereas other categories showed more variable or flat trajectories (Figure~\ref{by_func_fig1}).

 
At its best year, this suite of models achieves an average Spearman correlation of 0.38. When we perform ridge regression, this correlation jumps substantially, leading to the expected boost in performance and continuing to underscore the importance of experimental labels. Even training on 10\% of the dataset can increase performance to 0.52, and to around 0.675 when training on 80\% of the dataset (Figure~\ref{rand}A). 

Both the \textit{Contiguous} (Figure~\ref{rand}B) or \textit{Modulo} (Figure~\ref{modulo_fig2b}) data splitting schemes show that performance improves consistently with the inclusion of more labeled training groups. However, at any given group count, model performance does not show a consistent trend with respect to pre-training year. 

Finally, we zoom-in on the \(\beta\)-Lactamase datasets to test the hypothesis that, for this well-characterized protein family with abundant data, scaling laws may begin to emerge. Indeed, we see that the unsupervised predictions of fitness for all four datasets show improvements in correlation with most AMPLIFY models, beginning with a weak correlation of around 0.25 and improving to correlations more than 0.6 (Figure~\ref{BLAT}). When applying ridge regression in a semi-supervised setup—training on a single dataset and testing on others—we observe a stable correlation around 0.6 across all AMPLIFY models (lightest blue line in Figure~\ref{BLAT}). Notably, this line intersects the unsupervised curve around 2020, indicating that training on just one experimental dataset can match the zero-shot performance of much larger models trained on a decade of UniRef100 data. As additional datasets are incorporated into training, performance improves monotonically, with two- and three-dataset models outperforming all unsupervised baselines.


\subsection{Future Work}
We plan on extending this analysis to other families of pLMs as well as exploring other methods for semi-supervised learning. We hope to train a suite of models on both year- and clustering-splits of UniRef. We will also test additional splits between datasets to understand scaling laws in transfer learning. We will train on chosen datasets (i.e. prokaryotic, or a specific protein class, etc) and test on the remaining datasets in ProteinGym. We also hope to randomly split ProteinGym between datasets and train on a small number of randomly selected datasets and test on all others (i.e. train on 10 datasets and test on remaining 210+; train on 100 datasets and test on remaining 110+) to see how performance is balanced between large-scale experimental data and LLMs.

\section{Conclusion}
Our findings suggest that for the protein function prediction task, biological data scaling does not yet follow a simple, monotonic trend. Even as pretraining datasets grow, model performance remains sensitive to data composition and has yet to saturate. Experimental labels remain essential for boosting predictive power, and targeted benchmarks like \(\beta\)-Lactamase highlight where scaling benefits begin to emerge. Continued exploration of data scaling---across families, tasks, and learning paradigms---is needed to guide the next generation of biological language models. 

\section*{Data and Code Availability}
All supervised experiments in this work were conducted using the Substitution DMS dataset from the \href{https://github.com/OATML-Markslab/ProteinGym}{ProteinGym benchmark}. Specifically, we used \href{https://github.com/OATML-Markslab/ProteinGym/blob/bb685245f8f4bb95a5b54b470396c23826cd6284/reference_files/DMS_substitutions.csv}{DMS\_substitutions.csv}
 (protein-level metadata) and \href{https://github.com/OATML-Markslab/ProteinGym/blob/bb685245f8f4bb95a5b54b470396c23826cd6284/benchmarks/DMS_zero_shot/substitutions/Spearman/DMS_substitutions_Spearman_DMS_level.csv}{DMS\_substitutions\_Spearman\_DMS\_level.csv}
 (unsupervised model performance) from the official ProteinGym repository. Instructions for accessing the full dataset are available on the \href{https://github.com/OATML-Markslab/ProteinGym/tree/main?tab=readme-ov-file#resources}{ProteinGym Resources page}
.

All training, inference, analysis, and visualization code is available on our \href{https://github.com/Align-to-Innovate/data-saturation-and-scaling}{GitHub}.
Pretrained AMPLIFY model checkpoints for all years (2011–2024) can be found at:
\url{https://huggingface.co/chandar-lab/AMPLIFY_120M/tree/AMPLIFY_120M_}\texttt{<YEAR>} (replace \texttt{<YEAR>} with the desired year).

\section*{Impact Statement}

Work presented in this paper impacts the biological machine learning community in understanding the scaling laws underpinning biological data. Because we believe that current public data repositories do not saturate biological data for the task of predicting results of deep mutational scanning experiments, we hope this motivates more scientists to pursue data acquisition and curation. We also hope that the field moves towards quantifying existing data in these repositories \& creating better metrics for summarizing complex sequence data.  
\section*{Acknowledgments}
Thank you to the GenBio workshop reviewers and Alan Amin, Pascal Notin, Sarah Gurev, and Lood van Niekerk for helpful comments on the manuscript. We would like to thank the Align team for their support and contributions to this work. This work was supported by The Align Foundation, which receives philanthropic funding in part from Griffin Catalyst.

\bibliography{scaling_laws_bio}
\bibliographystyle{icml2025}

\newpage
\appendix
\onecolumn
\setcounter{figure}{0}
\renewcommand{\thefigure}{S\arabic{figure}}
\section{Appendix}

\setlength{\tabcolsep}{10pt} 
\renewcommand{\arraystretch}{1.3} 
\begin{table}[h!]
\centering
\begin{tabular}{|c|r|r|r|r|}
\hline
\textbf{Year} & \textbf{UniProt} & \textbf{UniRef100} & \textbf{UniRef90} & \textbf{UniRef50} \\
\hline
2011 & 13069501 & 11659891 & 7623063 & 3653743 \\
2012 & 19968488 & 15688962 & 9843844 & 4606913 \\
2013 & 29805788 & 20491136 & 12880369 & 6412887 \\
2014 & 52159208 & 33613081 & 20200107 & 9370012 \\
2015 & 89998523 & 50371270 & 28628106 & 11992242 \\
2016 & 60268458 & 72946704 & 39362473 & 16038089 \\
2017 & 74265355 & 94756963 & 49122202 & 20083468 \\
2018 & 108184003 & 133853533 & 69029793 & 30071646 \\
2019 & 140253338 & 172327164 & 87296736 & 32474829 \\
2020 & 178316438 & 216491817 & 107153647 & 39232797 \\
2021 & 208365010 & 262115656 & 130661074 & 49127834 \\
2022 & 230895644 & 297827854 & 144113457 & 51333317 \\
2023 & 246440937 & 342650445 & 166459614 & 59142917 \\
2024 & 250322721 & 390790959 & 184520054 & 63849054 \\
2025 & 253206171 & 453950711 & 204806910 & 69290910 \\
\hline
\end{tabular}
\caption{Counts of UniProt and UniRef entries by year.}
\label{tab:uniref-years}
\end{table}

\begin{figure}[ht]
\vskip 0.2in
\begin{center}
\centerline{\includegraphics[width=\columnwidth]{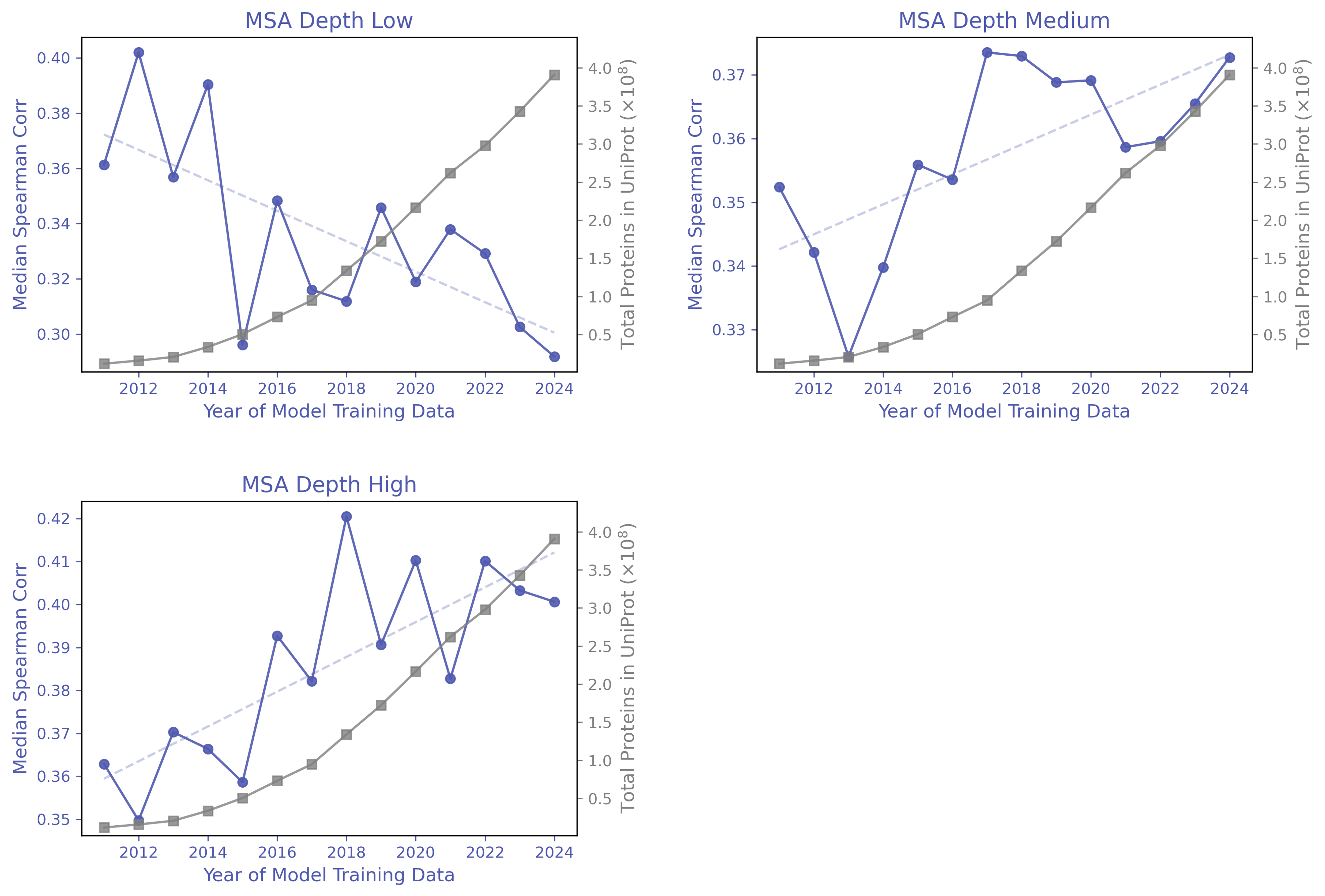}}
\caption{Figure~\ref{by_depth_fig1} replicates Figure~\ref{pg_correlation_by_year}, partitioned by MSA depth, as represented as Neff/L from ProteinGym. Proteins with Low MSA depth generally get worse with later timepoints while proteins with Medium and High MSA depth generally improve. MSA Neff/L categories were distributed as follows: Medium (106), High (72), and Low (35)}
\label{by_depth_fig1}
\end{center}
\vskip -0.2in
\end{figure}

\begin{figure}[ht]
\vskip 0.2in
\begin{center}
\centerline{\includegraphics[width=\columnwidth]{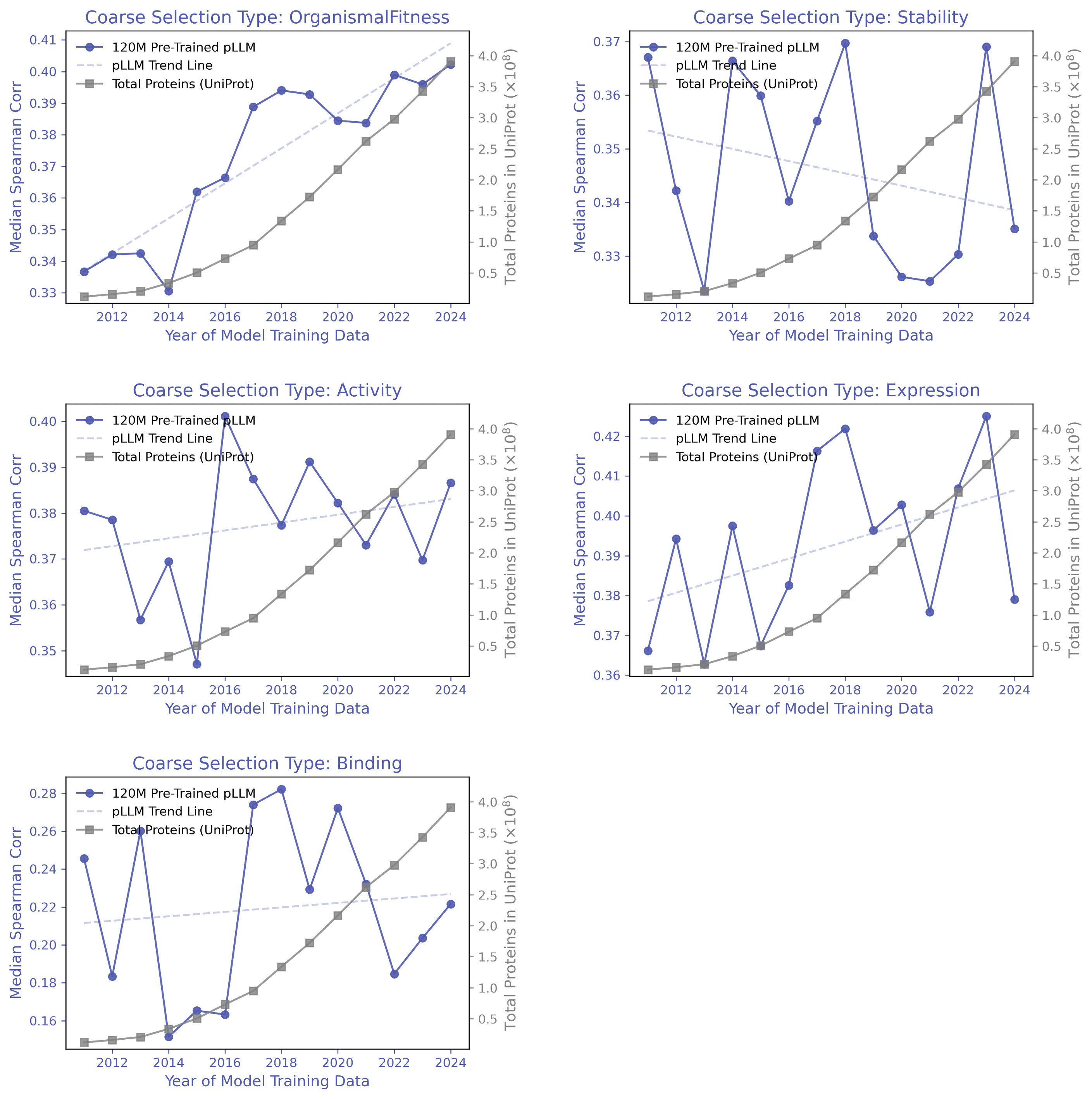}}
\caption{Figure~\ref{by_func_fig1} replicates Figure~\ref{pg_correlation_by_year}, partitioned by coarse selection type, as described in ProteinGym. Organismal Fitness has a steadily increasing trend whereas all other categories fluctuate. Functional categories were represented by: OrganismalFitness (73), Stability (66), Activity (43), Expression (18), and Binding (13).}
\label{by_func_fig1}
\end{center}
\vskip -0.2in
\end{figure}

\begin{figure}[ht]
\vskip 0.2in
\begin{center}
\centerline{\includegraphics[width=\columnwidth]{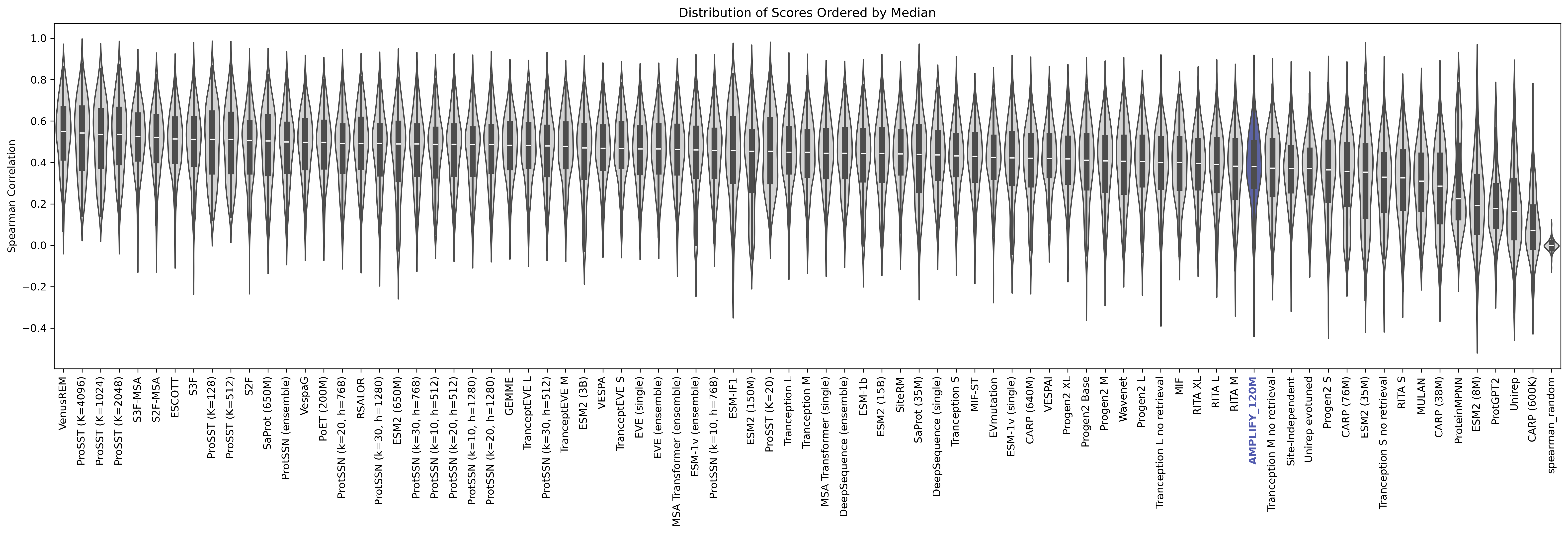}}
\caption{AMPLIFY120M trained on UniRef100 from 2024 has similar model performance to RITA\cite{hesslow2022rita}, Tranception\cite{notin2022tranception}, and the smaller ESM2\cite{lin2023evolutionary} and ProGen\cite{nijkamp2023progen2} models.}
\label{all_in_PG}
\end{center}
\vskip -0.2in
\end{figure}

\begin{figure}[ht]
\vskip 0.2in
\begin{center}
\centerline{\includegraphics[width=\columnwidth]{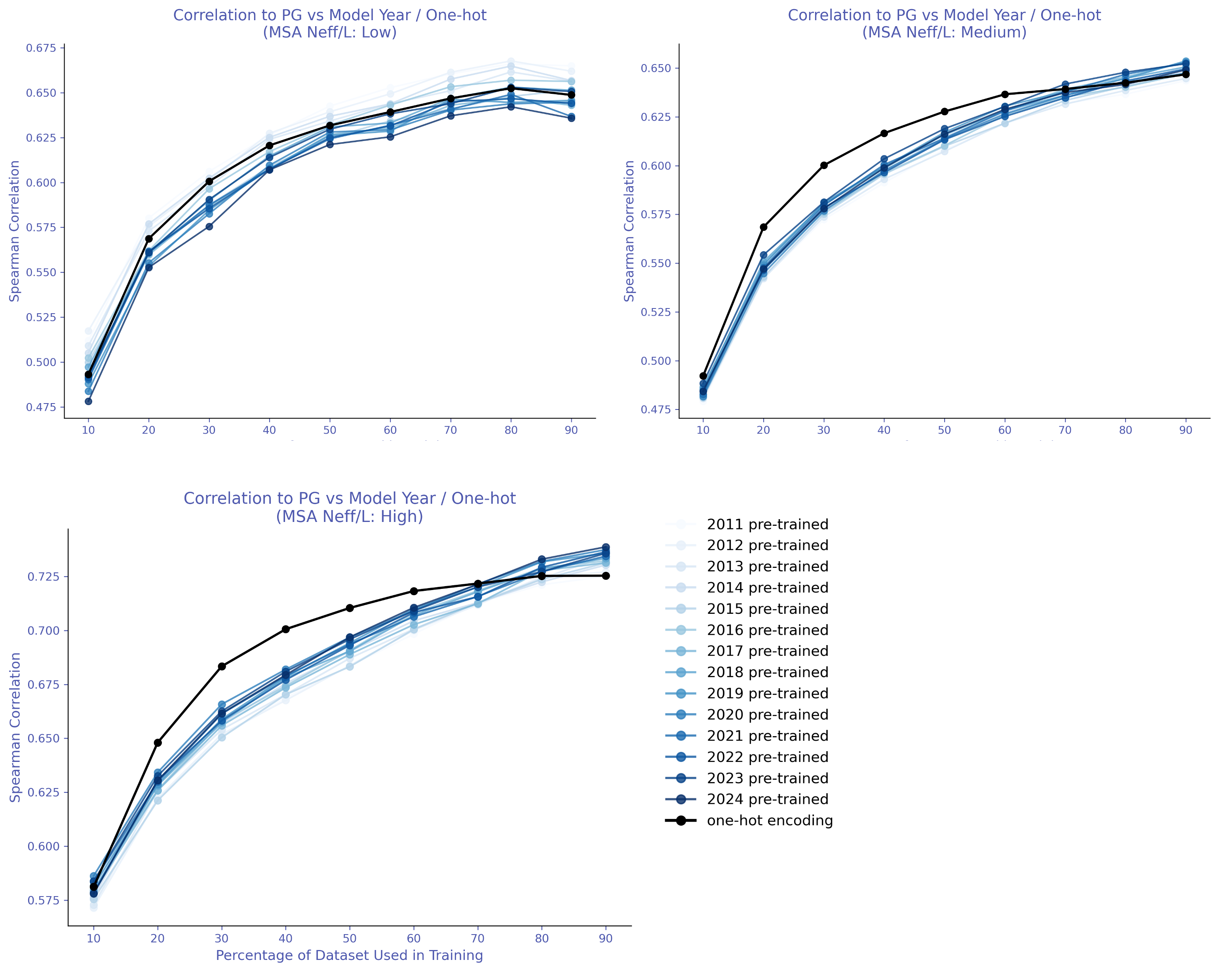}}
\caption{Figure~\ref{by_depth_fig2} replicates Figure~\ref{rand}A, partitioned by by MSA depth, as represented as Neff/L from ProteinGym. Proteins with Low MSA depth exhibit better performance with models trained on earlier timepoints of UniRef100. Whereas proteins with Medium and High MSA depth do not show as large of a range.}
\label{by_depth_fig2}
\end{center}
\vskip -0.2in
\end{figure}

\begin{figure}[ht]
\vskip 0.2in
\begin{center}
\centerline{\includegraphics[width=\columnwidth]{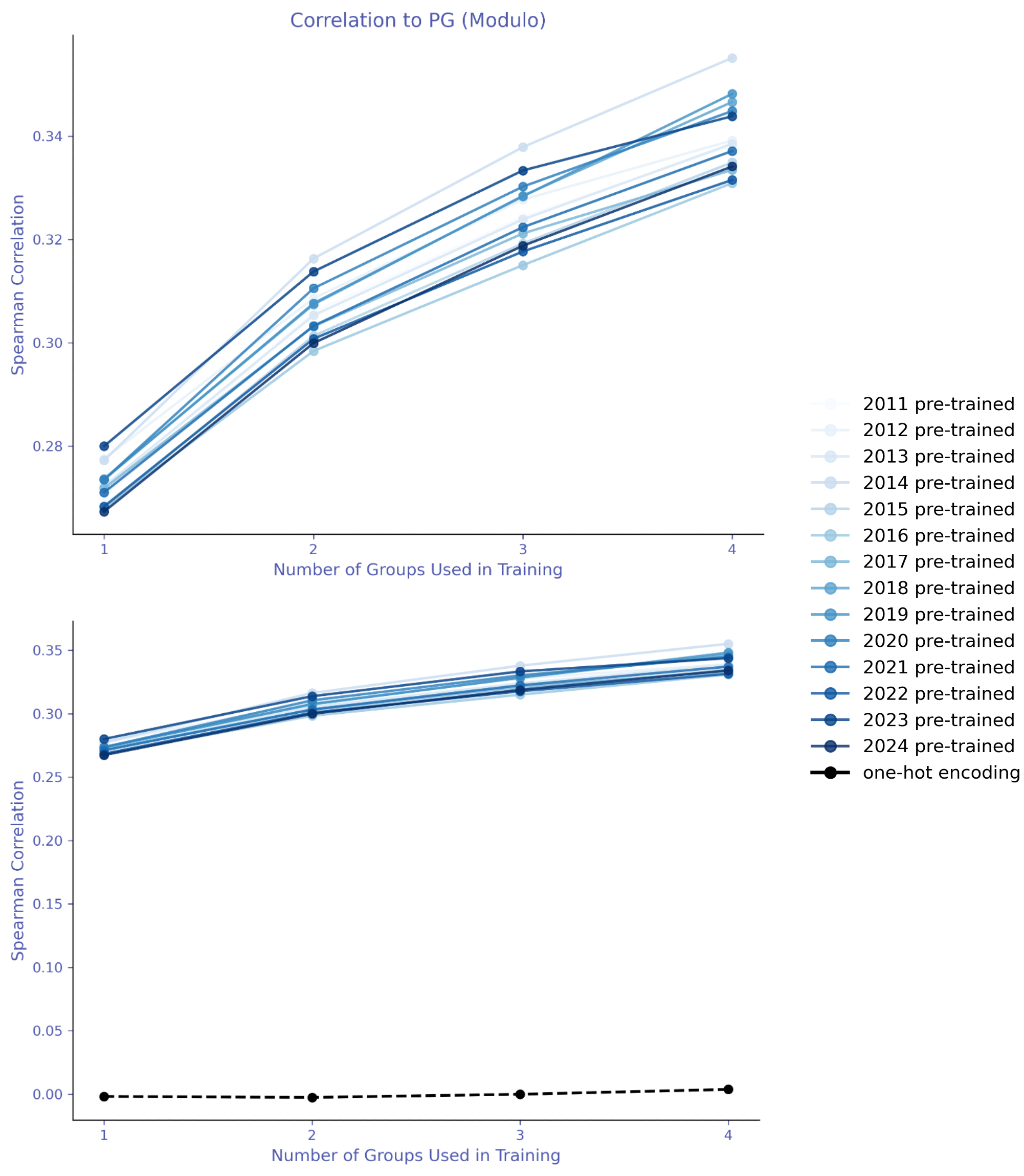}}
\caption{This is a similar plot as Figure 2B but with the Modulo train/test split. The top plot shows the results without one-hot plotted and the bottom plot shows the results with one-hot plotted. One-hot encodings provide almost no information to the model when splitting in this non-random way and the correlation hovers around 0 regardless of how much labeled data is used.}
\label{modulo_fig2b}
\end{center}
\vskip -0.2in
\end{figure}

\begin{figure}[ht]
\vskip 0.2in
\begin{center}
\centerline{\includegraphics[width=\columnwidth]{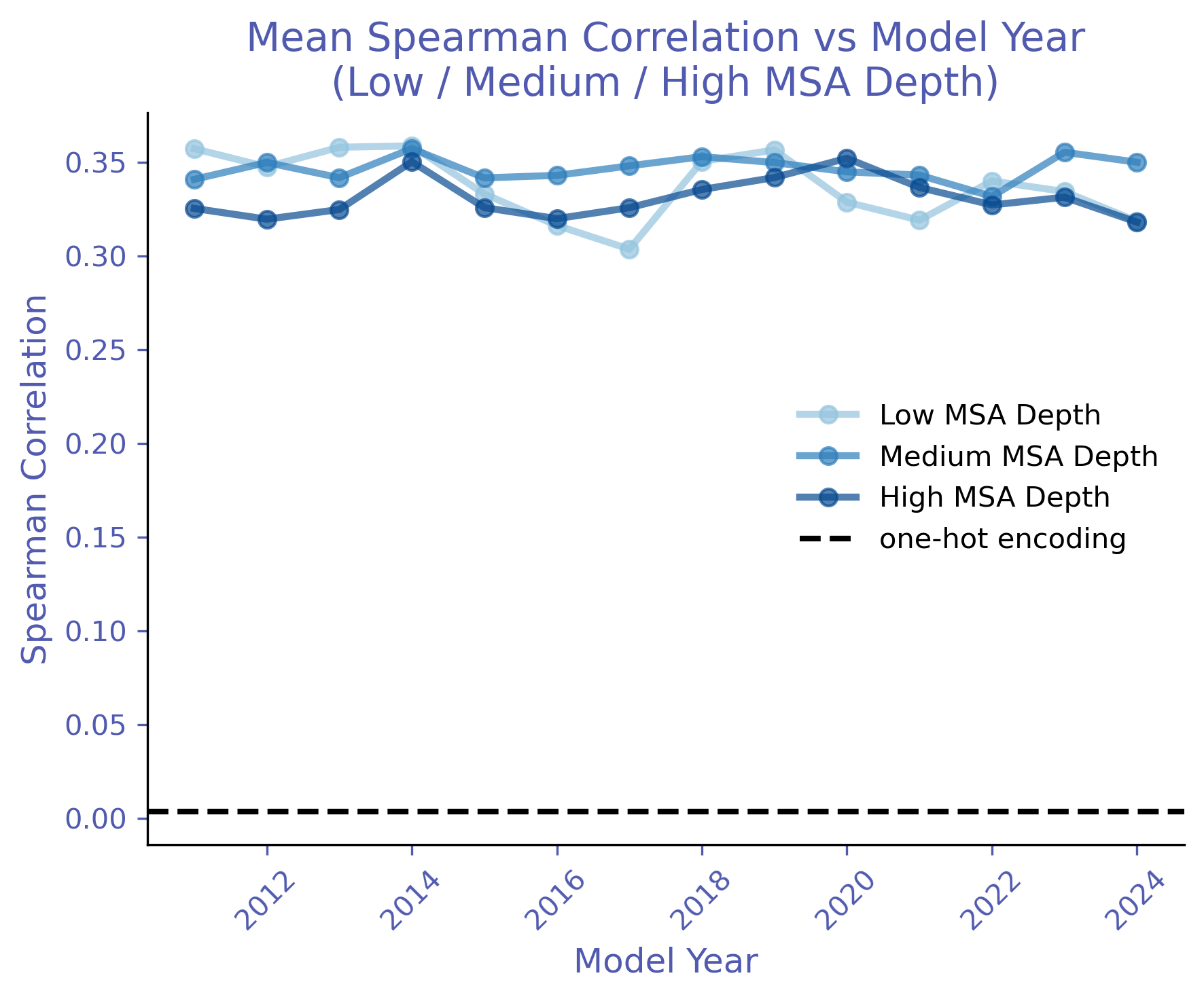}}
\caption{When looking at the Modulo train/test splits by MSA depth, we see that the trend is similar between them and relatively flat.}
\label{modulo_by_MSA_depth}
\end{center}
\vskip -0.2in
\end{figure}

\begin{figure}[ht]
\vskip 0.2in
\begin{center}
\centerline{\includegraphics[width=\columnwidth]{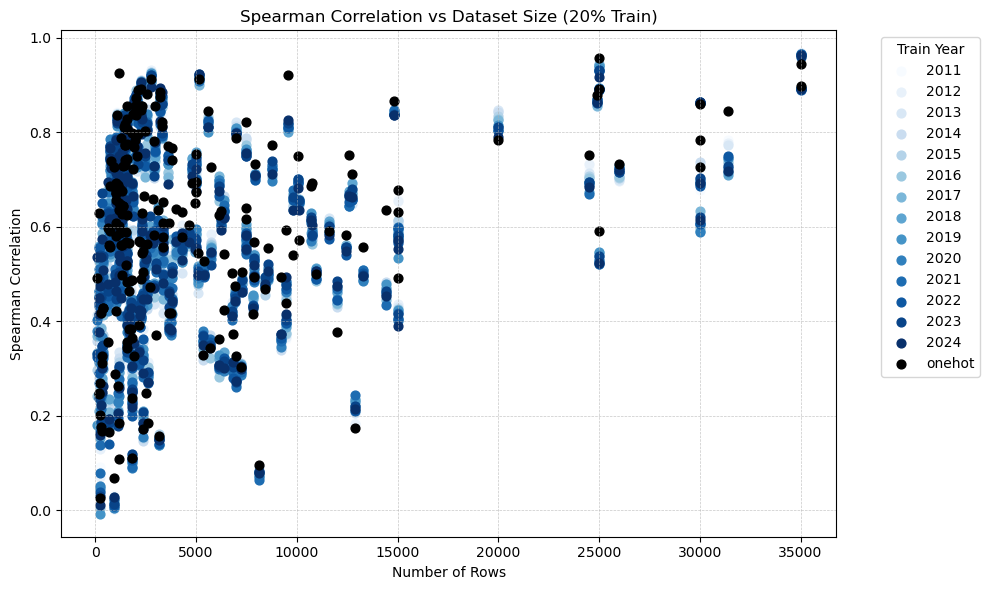}}
\caption{The number of mutants included in the dataset impacts the spread of spearman correlation between the model predictions and the experimental values with random split. With relatively smaller dateset sizes, the spearman correlations range from 0 to close to 1. Whereas with larger dataset sizes (\>15,000 mutants measured), the model accuracy is always above 0.5. This is only for training on a random split of 20\% of the data and we see a similar trend across all random data splits.}
\label{random_by_rows}
\end{center}
\vskip -0.2in
\end{figure}

\begin{figure}[ht]
\vskip 0.2in
\begin{center}
\centerline{\includegraphics[width=\columnwidth]{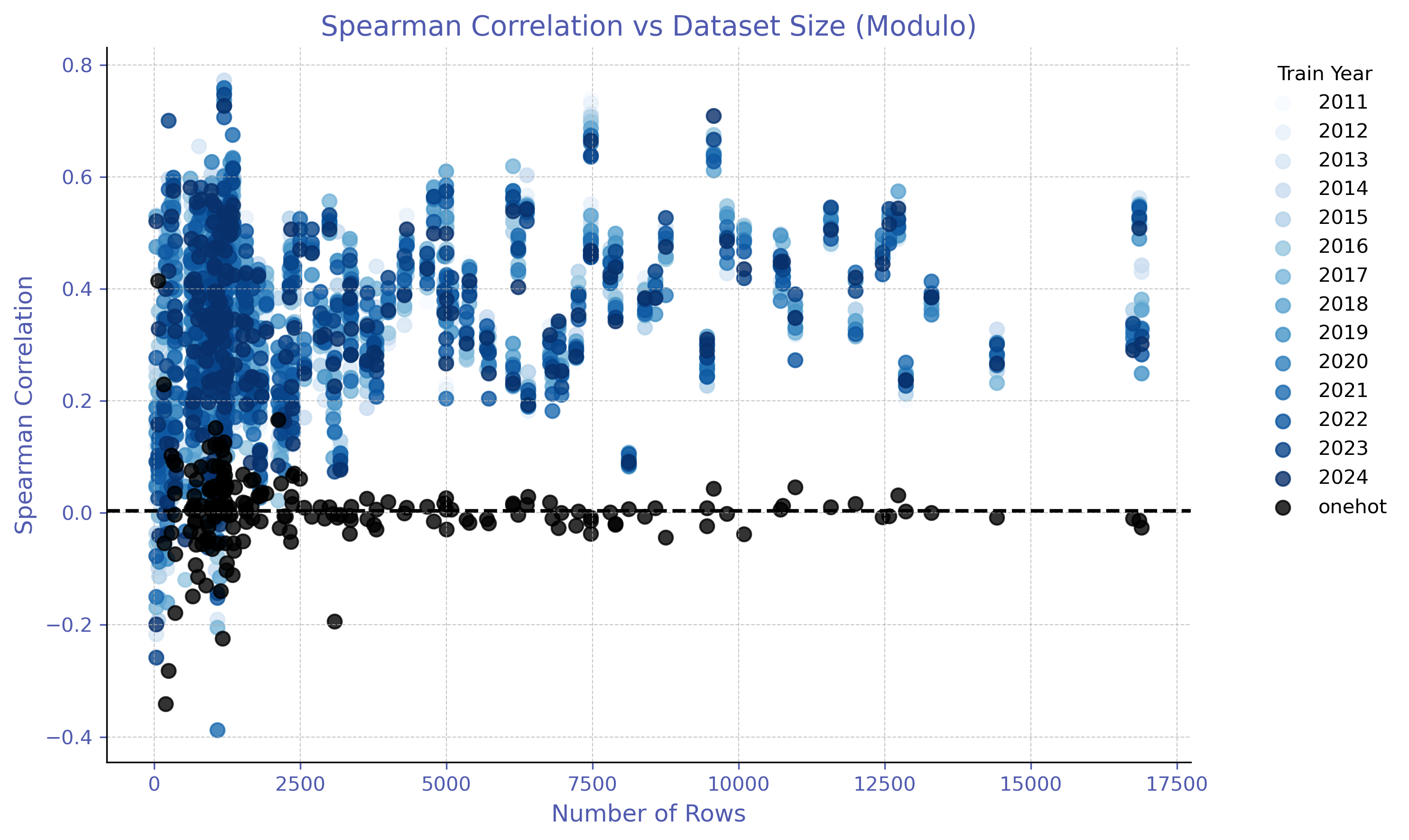}}
\caption{Identical Figure~\ref{random_by_rows} but with the Modulo train/test split instead of random. Generally we see a similar trend of more spread out data at lower number of mutations (rows). One-hot encoding, however, does look incredibly different and hovers around 0, with the mean correlation shown in the dashed line.}
\label{modulo_by_rows}
\end{center}
\vskip -0.2in
\end{figure}

\end{document}